\documentstyle[12pt,epsf]{article}
\def\al{\alpha}
\def\be{\beta}
\def\ga{\gamma}
\def\de{\delta}
\def\ep{\epsilon}

\def\et{\eta}

\def\ta{\tau}

\def\ps{\psi}

\def\Ga{\Gamma}
\def\De{\Delta}

\def\Up{\Upsilon}

\def\fr#1#2{{{#1} \over {#2}}}

\def\bra#1{\langle{#1}|}
\def\ket#1{|{#1}\rangle}

\def\half{{\textstyle{1\over 2}}}
\def\frac#1#2{{\textstyle{{#1}\over {#2}}}}

\def\lsim{\mathrel{\rlap{\lower4pt\hbox{\hskip1pt$\sim$}}
    \raise1pt\hbox{$<$}}}
\def\gsim{\mathrel{\rlap{\lower4pt\hbox{\hskip1pt$\sim$}}
    \raise1pt\hbox{$>$}}}
\def\sqr#1#2{{\vcenter{\vbox{\hrule height.#2pt
         \hbox{\vrule width.#2pt height#1pt \kern#1pt
         \vrule width.#2pt}
         \hrule height.#2pt}}}}

\def\Re{\hbox{Re}\,}
\def\Im{\hbox{Im}\,}

\def\z{{\bf\hat z}}

\newcommand{\beq}{\begin{equation}}
\newcommand{\eeq}{\end{equation}}
\newcommand{\bea}{\begin{eqnarray}}
\newcommand{\eea}{\end{eqnarray}}
\newcommand{\rf}[1]{(\ref{#1})}
\newcommand{\ct}[1]{\cite{#1}}

\renewenvironment{thebibliography}[1]
 { \rm
   \begin{list}{\arabic{enumi}.}
    {\usecounter{enumi} \setlength{\parsep}{0pt}
     \setlength{\itemsep}{3pt} \settowidth{\labelwidth}{#1.}
     \sloppy
    }}{\end{list}}
 
\begin{document}
\titlepage
 
\begin{flushright}
{IUHET 334\\}
{May 1996\\}
\end{flushright}
\vglue 1cm
	    
\begin{center}
{{\bf BOUNDING CPT VIOLATION IN THE NEUTRAL--B SYSTEM 
\\}
\vglue 1.0cm
{V. Alan Kosteleck\'y and R. Van Kooten\\} 
\bigskip
{\it Physics Department\\}
\medskip
{\it Indiana University\\}
\medskip
{\it Bloomington, IN 47405, U.S.A.\\}
 
\vglue 0.8cm
}
\vglue 0.3cm
 
\end{center}
 
{\rightskip=3pc\leftskip=3pc\noindent
The feasibility of placing bounds on CPT violation
from experiments with neutral-$B$ mesons is examined.
We consider situations with uncorrelated mesons 
and ones with either unboosted or boosted correlated mesons. 
Analytical expressions valid for 
small T- and CPT-violating parameters are presented 
for time-dependent and time-integrated decay rates,
and various relevant asymmetries are derived.
We use Monte-Carlo simulations to model 
experimental conditions for a plausible range
of CPT-violating parameters.
The treatment uses realistic data 
incorporating background effects, resolutions, and acceptances
for typical detectors at LEP, CESR, and the future $B$ factories.   
Presently, 
there are no bounds on CPT violation in the $B$ system.
We demonstrate that limits of order 10\% on CPT violation 
can be obtained from data already extant,
and we determine the CPT reach attainable within the next few years.

}
 
\vskip 1truein
\centerline{\it Accepted for publication in Physical Review D}
 
\vfill
\newpage
 
\baselineskip=20pt

{\bf\noindent I. Introduction}
\vglue 0.4cm

With the advent of experiments
capable of reconstructing relatively large numbers
of decays of neutral $B$ mesons,
a new window has opened for high-precision tests 
using the interferometric nature of the neutral $B$ system
in analogy to the neutral-kaon system. 
A particularly interesting possibility 
is testing CPT invariance,
which is believed to be a fundamental symmetry
of local relativistic particle field theories
\cite{s1,s2,s3,s4,s5,s6,s7}
and therefore also of the standard model.
The possibility that CPT invariance
might be spontaneously violated in string theory
at observable levels via a mechanism arising
as an indirect effect of string nonlocality 
has been suggested 
in refs.\ \cite{kp1,kp11,kp2}.
CPT violation might also arise 
if quantum mechanics is modified by gravity,
\cite{sh,dp,rw},
including perhaps in the context of string theory
with unconventional quantum mechanics
\cite{emn}.

At present,
the tightest bounds on CPT violation
come from observations in the neutral-kaon system 
\cite{c1,c2,c3,tr}.
However,
experimental bounds could also be obtained using 
other neutral-meson systems.
This possibility is of interest
because the magnitudes of CPT violation
could be different in the various systems.
For example,
in the string scenario with spontaneous CPT violation,
effects of different magnitude
could appear at levels accessible to future 
and perhaps even present experiments
not only in the $K$ system but also in the $B$ and $D$ systems
\cite{kp2}.
It has recently been demonstrated that
CPT signals can in principle be separated from other effects 
in the neutral-$B$ system using $B$ factories
\cite{ck1}
and in the neutral-$D$ system using both
correlated and uncorrelated decays
\cite{ck2}.
These analyses were purely theoretical,
disregarding background effects
and detector acceptances.
Otherwise,
with a few notable exceptions
\cite{ks,zx},
the issue of CPT violation in the $B$ system
has received relatively little attention
in the literature.
At present,
there are no experimental limits
on CPT violation in either the $B$ or the $D$ systems.

Several types of experiments collecting
$B$ events are presently underway or
under construction.
Close to $10^6$ reconstructed 
uncorrelated $B$ events are already available
from each of the LEP collaborations at CERN
\cite{lp}.  
Large numbers of uncorrelated $B$ events have also been collected
at the Tevatron
\cite{tev}. 
A symmetric $B$ factory using CESR at Cornell
has produced relatively large numbers of
correlated $B^0_d$-$\overline{B^0_d}$ pairs
from the decay of the $\Up (4S)$ resonance,
with about $2.6 \times 10^6$ reconstructed $\Up (4S)$ decay events
obtained in CLEO
\cite{cl1}.
Asymmetric $B$ factories,
which generate boosted correlated $B$ pairs,
are under construction at SLAC and KEK.
The corresponding detectors BaBar and BELLE
should provide approximately $10^7$ reconstructed $B$ pairs
in a year at design luminosity
\cite{bf1}.

In the present work,
we investigate the practicability of
extracting bounds on CPT violation from
a variety of realistic experimental situations.
We obtain analytical expressions for time-dependent
and time-integrated decay rates
that hold under general circumstances 
for uncorrelated,
correlated unboosted,
and correlated boosted $B$ mesons.
In each case,
we use Monte-Carlo simulations to 
provide a measure of the level at which CPT violation
can be bounded.
For definiteness, we perform simulations assuming a
typical LEP detector (OPAL) at CERN
\cite{opal1},
the detector CLEO at CESR
\cite{cleo1},
and the detector BaBar at SLAC
\cite{tdrb}.

Our basic notation and conventions are presented 
in Sec.\ II,
along with an outline of the procedure we adopt
for the Monte-Carlo simulations.
We begin the detailed analysis in Sec.\ III,
which treats the case of uncorrelated $B$ mesons.
The theory for this situation is in Sec.\ IIIA,
while the experimental simulation 
involving the detector OPAL at CERN is in Sec.\ IIIB.
Section IV discusses the case with correlated $B$ mesons.
The theoretical analysis is in Sec.\ IVA, 
and the experimental simulations for the unboosted (CLEO at CESR) 
and the boosted (BaBar at SLAC) situations 
are in Secs.\ IVB and IVC, 
respectively.
Section V contains a summary.

\vglue 0.6cm
{\bf\noindent II. Preliminaries}
\vglue 0.4cm

This section contains basic definitions
underlying our analyses
and an outline of our simulation procedures.
Most of our conventions
are related to canonical ones for the kaon system
and are discussed in more detail in 
refs.\ \ct{kp2,ck1}.
Our analysis assumes throughout that any CP violation is small,
which implies small T and CPT violation,
and it disregards terms that are higher-order
in small quantities.

The effective hamiltonian 
for the $B^0$-$\overline{B^0}$ system 
has eigenvectors given by
\bea
\ket{B_S} & = & 
[(1 + \ep_B + \de_B)\ket{B^0}
+(1 - \ep_B - \de_B)\ket{\overline{B^0}}]/{\sqrt 2}  
\quad , \nonumber\\
\ket{B_L} & = & 
[(1 + \ep_B - \de_B)\ket{B^0}
-(1 - \ep_B + \de_B)\ket{\overline{B^0}}]/{\sqrt 2} 
\quad . 
\label{iia}
\eea
The quantities $\ep_B$ and $\de_B$
are complex and parametrize CP violation.
The former measures indirect T violation,
while the latter measures indirect CPT violation.
The analogous parameters 
$\ep_K$ and $\de_K$ for the $K$ system
are defined by Eq.\ \rf{iia} 
but with the replacement $B\to K$.

The time evolution of the physical states
$B_S$, $B_L$ 
is governed by the corresponding eigenvalues of
the effective hamiltonian:
\beq
\ket{B_S(t)}=\exp ({-im_St-\ga_St/2}) \ket{B_S}
\quad , \quad
\ket{B_L(t)}=\exp ({-im_Lt-\ga_Lt/2}) \ket{B_L}
\quad ,
\label{iib}
\eeq
where 
the physical masses are $m_S$, $m_L$ 
and the decay rates are $\ga_S$, $\ga_L$.
It is convenient to introduce definitions 
for certain frequently occurring combinations
of these quantities.
We set
\bea
m = m_S + m_L
\quad &,& \qquad
\ga=\ga_{S}+\ga_{L}
\quad , \qquad
\nonumber\\
\De m = m_L - m_S
\quad &,& \qquad
\De \ga = \ga_S - \ga_L
\quad , \qquad
\nonumber\\
a^2 = \De m^2 + \De \ga^2/4
\quad &,& \qquad
b^2 = \De m^2 + \ga^2/4
\quad , \qquad
\nonumber\\
x = \fr {2\De m}{\ga}
\quad &,& \qquad
y = \fr{\De\ga}{\ga}
\quad .
\label{iic}
\eea

Experimental data 
are available for some of these quantities
\cite{pdt}, 
specifically for the $B^0_d$ meson
\cite{foot}.
The mass difference is
$|\De m_d | = (3.4 \pm 0.3) \times 10^{-13}$ GeV,
and the mean lifetime is 
$\bar\ta_{B^0} = (1.50\pm 0.11)\times 10^{-12}$ s.
These values determine the
magnitude of the mixing parameter $x$ as
$ |x_d| \ge 0.71 \pm 0.06$.
We introduce the inequality here
to allow for nonzero CPT violation,
which affects the standard analysis 
\cite{ks}
and increases the numerical value of $|x_d|$.

In contrast,
no experimental limit exists on
the scaled rate difference $y$.
It is theoretically plausible to take the magnitude of $y$ as  
$|y| \simeq O(10^{-2}) \ll |x|$.
This estimate is based on 
the box diagram in perturbation theory,
where the dominant intermediate states
are the top and charm quarks.
It is likely to be a good approximation
since short-distance effects are larger than dispersive ones
in the neutral-$B$ system.
More details about this analysis and references to
the literature can be found in,
for example,
the reviews
\cite{lc,pf,pt,ib}.

For the purposes of the present work,
we focus on two types of $B^0$-decay modes:
the semileptonic modes $B^0 \to D^{(*)} \ell \nu$
with $\ell$ being $e$ or $\mu$ only,
and the mode $B^0\rightarrow J/\ps K_S$.
These are particularly attractive from the experimental viewpoint.
In principle,
our CPT analyses could be further enhanced 
by grouping together certain special $B$ decays,
called \it semileptonic-type \rm decays
\cite{ck1}.
These include the standard semileptonic decays above, 
together with a subset of other modes
$B^0 \rightarrow f$
for which there is no lowest-order weak process 
allowing a significant contamination of either 
$\overline{B^0} \rightarrow f$ 
or $B^0 \rightarrow \overline f$. 
Observed semileptonic-type decays include
\cite{pdt}
$B^0\rightarrow D^-D_s^+$, 
$B^0\rightarrow J/\ps K^+\pi^-$, 
$B^0\rightarrow J/\ps K^{*0}(892)$, 
$B^0\rightarrow \ps(2S)K^{*0}(892)$, 
and similar decays into excited states.
Observed modes other than those listed
incorporate a CKM-suppressed process
excluding them from the semileptonic-type class.
The theoretical analyses in the present work 
are general enough to allow
for this special grouping,
but for simplicity we disregard it in our Monte-Carlo simulations.
Incorporating it should result in some improvement 
over the bounds we obtain.

The analyses in the sections below 
involve transition amplitudes
for the decay into a final state $f$,
taken to be either a semileptonic-type state 
or a $J/\ps K^0$ (or conjugate) state.
Following standard procedure,
we disregard possible effects from penguin diagrams 
or other loop contributions
and parametrize these amplitudes as
\cite{lw,ba,td}
\bea
\bra{f}T\ket{B^0}
= F_f (1 - y_f)~~~~,~~ &
\bra{f}T\ket{\overline{B^0}} 
= x_f F_f (1 - y_f)
\quad , \nonumber \\
\bra {\overline f}T\ket {\overline{B^0}} = 
F_f^*(1 + y_f^*)~~~~,~~ &
\bra {\overline f}T\ket{B^0} 
= {\overline x_f^*} F_f^* (1 + y_f^*)
\quad .
\label{iid}
\eea
The independent complex quantities $x_f$, 
$\overline x_f$, $F_f$, and $y_f$ 
are assumed small in what follows.
Any possible violation of the
$\De B = \De Q$ rule is parametrized by
$x_f$ and $\overline x_f$,
which both vanish if the rule is exact.
Invariance under T implies 
$x_f$, $\overline x_f$, $F_f$, and $y_f$ are real.
Invariance under CPT implies
$x_f = \overline x_f$ and $y_f = 0$.
The quantity $y_f$ parametrizes 
direct CPT violation 
in the decay to $f$.

Since the observed states are $B_S$ and $B_L$,
it is useful to obtain the associated 
transition amplitudes analogous to those
in Eq.\ \rf{iid}.
Combining the definitions \rf{iia}, \rf{iid}
and keeping only terms to first order in small quantities
gives
\bea
\bra{f}T\ket{B_S}
& = & \frac 1 {\sqrt 2} F_f 
(1 + \ep_B + \de_B - y_f + x_f)
\quad , \nonumber \\
\bra{f}T\ket{B_L}
& = & \frac 1 {\sqrt 2} F_f
(1 + \ep_B - \de_B - y_f - x_f) 
\quad , \nonumber \\
\bra {\overline f}T\ket{B_S} 
& = & \frac 1 {\sqrt 2} F^*_f
(1 - \ep_B - \de_B + y^*_f + \overline x_f^*)
\quad , \nonumber \\
\bra {\overline f}T\ket{B_L} 
& = & - \frac 1 {\sqrt 2} F^*_f
(1 - \ep_B + \de_B + y^*_f - \overline x_f^*)
\quad .
\label{iie}
\eea

The theoretical analyses in the following sections
incorporate the possibility of indirect T and CPT violation  
through the parameters $\ep_B$ and $\de_B$
and allows for various sources of T and CPT violation 
through the parameters $F_f$, $x_f$, $\overline{x}_f$,
and $y_f$.
This level of generality has been kept for completeness.
However, 
it is probable that some of these parameters 
are much smaller than others.  
For example, if CPT breaking does occur, 
general theoretical prejudice suggests that 
the magnitude of any indirect CPT violation is likely
to be substantially larger than 
that of the direct CPT violation in 
any given channel.
This is because $\de_B$
is an interferometric parameter,
and moreover one that results
from the combined effects over all channels
of sources of CPT violation.
These ideas are supported by 
analysis in the string scenario with spontaneously broken CPT,
where quantities such as $\Re y_f$ are suppressed
by many orders of magnitude relative to $\de_B$
\cite{kp2}.
Note that 
in the standard purely phenomenological description
the parameters $\Re\de_B$ and $\Im\de_B$ are 
undetermined and independent.
In contrast,
in the string scenario they are 
determined in terms of other quantities in the theory
and are related through an additional constraint,
\beq
x\Im\de_B \pm y\Re\de_B = 0
\quad . 
\label{iif}
\eeq
For generality,
we do \it not \rm impose this condition in what follows.

A Monte-Carlo simulation of a realistic 
experimental situation allowing for the full
parameter range of all the available T- and
CPT-violating parameters produces unwieldy results.
For the illustrative purposes of the present work,
we have therefore chosen 
to restrict the multiplicity of available variables 
as much as possible compatible with
our goal of demonstrating the feasibility
of extracting CPT bounds from realistic experimental data.
In our simulations,
we have taken advantage of the
likely hierarchy in the magnitude of the various parameters,
as described above,
and have kept only quantities affecting
indirect T and CPT violation.
This restriction provides a meaningful
approximation for extracting CPT bounds
from experimental data.

As described in the Introduction,
the present work contains Monte-Carlo simulations 
of uncorrelated decays at a collider
and of correlated decays at both symmetric 
and asymmetric $B$ factories.
We obtain four-vectors for $B$ events from $Z^0$ decays
and $\Up(4S)$ decays using the 
JETSET7.3 program
\cite{jetset} 
that models jet fragmentation, 
particle decays, 
and final-state parton showers.
The full chain of detector simulation 
and event reconstruction is complex and unnecessary
for our present purposes, 
so instead we statistically smear
the four-vectors obtained from JETSET7.3 
according to the energy and momentum resolutions outlined
for the OPAL detector
\cite{opal1},
the CLEOII detector
\cite{cleo1}, and
the BaBar detector
\cite{tdrb}.
Detector geometric acceptances are incorporated
in all cases,
as well as typical reconstruction efficiencies
estimated from existing publications or projected values.
The possibility of measurement of various asymmetries 
as a function of proper decay time is considered, 
and typical resolutions of decay lengths 
and $B^0$-meson boosts are taken 
for each of the detectors mentioned.
Additional details of the simulations 
are described in the relevant sections below.

\vglue 0.6cm
{\bf\noindent III. Uncorrelated Systems}
\vglue 0.4cm

{\bf\noindent A. Theory}
\vglue 0.4cm

This subsection provides the theoretical derivation 
of time-dependent and time-integrated amplitudes and rates
for relevant decays of uncorrelated neutral $B$ mesons.

We begin by considering decays into semileptonic-type 
final states $f$.
Suppose an uncorrelated meson state $\ket{B^0}$ 
is produced at time $t=0$
and evolves to the state $\ket{B(t)}$ after a time $t$
measured in the rest frame of the meson. 
The analogous quantities for an antimeson
are $\ket{\overline{B^0}}$ at $t=0$
and $\ket{\overline{B}(t)}$ at time $t$. 

To obtain the time-dependent decay amplitudes,
we proceed by inverting Eq.\ \rf{iia} to find the
states $\ket{B^0}$ and $\ket{\overline{B^0}}$
and incorporating the time evolution via Eq.\ \rf{iib}.
Using Eq.\ \rf{iid},
some algebra then leads to the 
following time-dependent decay probabilities:
\bea
P_f(t) 
& \equiv &
|\bra{f}T\ket{B(t)}|^2
\nonumber\\
&=&
\frac 1 4 |F_f|^2
\bigg(
(1 + 4\Re\de_B - 2\Re y_f + 2 \Re x_f)\exp(-\ga_S t)
\nonumber\\
&&\quad
+ (1 - 4\Re\de_B - 2\Re y_f - 2 \Re x_f)\exp(-\ga_L t)
\nonumber\\
&&\quad
+ 2 \bigl[ (1 - 2 \Re y_f) \cos \De m t
          - (4 \Im \de_B + 2 \Im x_f) \sin \De m t 
	  \bigr] \exp(-\ga t/2) 
\bigg) 
\quad , \nonumber\\
\overline{P}_{\overline f} (t)
& \equiv &
|\bra{\overline f}T\ket{\overline B(t)}|^2
\nonumber\\
&=& 
P_f(\de_B \rightarrow -\de_B,
y_f \rightarrow -y_f, 
x_f \rightarrow \overline{x}_f^*)
\quad , \nonumber\\
P_{\overline f} (t) 
& \equiv &
|\bra{\overline f}T\ket{B(t)}|^2
\nonumber\\
&=& 
\frac 1 4 |F_f|^2
\bigg(
(1 - 4\Re\ep_B + 2\Re y_f + 2 \Re \overline{x}_f)\exp(-\ga_S t)
\nonumber\\
&&\quad
+(1 - 4\Re\ep_B + 2\Re y_f - 2 \Re \overline{x}_f)\exp(-\ga_L t)
\nonumber\\
&&\quad
- 2 \bigl[ (1 - 4 \Re \ep_B + 2 \Re y_f ) \cos \De m t
          + 2 \Im \overline{x}_f \sin \De m t 
	  \bigr] \exp(-\ga t/2) 
\bigg) 
\quad , \nonumber\\
\overline{P}_f(t) 
& \equiv &
|\bra{f}T\ket{\overline B(t)}|^2
\nonumber\\
&=& 
P_{\overline f}
(\ep_B \rightarrow -\ep_B,
y_f \rightarrow -y_f, 
\overline{x}_f \rightarrow x_f^*)
\quad .
\label{iiia}
\eea
The decay probabilities 
$\overline{P}_{\overline f}$ and $\overline{P}_f$ 
are obtained by substituting as indicated
in the expressions for $P_f$ and $P_{\overline f}$.
These probabilities are used in subsection IIIB
as input for our Monte-Carlo simulations.

Intuition about the physical content of the
decay probabilities can be obtained by considering
the fully time-integrated rates and constructing
various asymmetries.
Explicitly, 
the time-integrated decay rates are:
\bea
R_f 
& \equiv & 
\int_{0}^{\infty} dt P_f 
\nonumber \\
& = & \frac 1 4 |F_f|^2 
\bigg( \ga
\Big[ \fr{1}{\ga_S \ga_L} +\fr{1}{b^2} \Big] (1 - 2\Re y_f) 
\nonumber \\
& & \qquad
- \fr{2 \De \ga}{\ga_S \ga_L} (2\Re \de_B + \Re x_f) 
- \fr {4 \De m}{b^2} (2\Im \de_B + \Im x_f)
\bigg)
\quad , \nonumber\\
\overline{R}_{\overline f} 
& \equiv & 
\int_{0}^{\infty} dt 
\overline{P}_{\overline f} 
= R_f(\de_B \rightarrow -\de_B,
y_f \rightarrow -y_f, 
x_f \rightarrow \overline{x}_f^*)
\quad , \nonumber\\
R_{\overline f} 
& \equiv & 
\int_{0}^{\infty} dt 
P_{\overline f}  
\nonumber \\
& = & \frac 1 4 |F_f|^2 
\bigg( \ga
\Big[ \fr{1}{\ga_S \ga_L} -\fr{1}{b^2} \Big] 
(1 - 4\Re \ep_B + 2 \Re y_f) 
\nonumber \\
& & \qquad
- \fr{2 \De \ga}{\ga_S \ga_L} \Re \overline x_f 
- \fr {4 \De m}{b^2} \Im \overline x_f 
\bigg)
\quad , \nonumber\\
\overline{R}_{f} 
& \equiv & 
\int_{0}^{\infty} dt 
\overline{P}_f 
= R_{\overline f}(\ep_B \rightarrow -\ep_B,
y_f \rightarrow -y_f, 
\overline{x}_f \rightarrow x_f^*)
\quad .
\label{iiib}
\eea

Two interesting independent asymmetries 
can be defined from these rates.
One is
\bea
A_f & \equiv & \fr{R_f - \overline R_{\overline f}}
{R_f + \overline R_{\overline f}}
\nonumber \\
& = & 
-2 \Re y_f - \fr{1}{\ga (b^2 + \ga_S \ga_L)}
\Bigl[\De \ga b^2(4 \Re \de_B +\Re (x_f - \overline x_f))
\nonumber \\
& & \qquad\qquad\qquad\qquad
+ 2 \De m \ga_S \ga_L (4 \Im \de_B + \Im (x_f + 
\overline x_f))\Bigr]
\quad .
\label{iiic}
\eea
The other is 
\bea
A_f^{\prime} &\equiv &\fr{\overline R_f - R_{\overline f}}
{\overline R_f + R_{\overline f}} 
\nonumber \\
& = &
4\Re\ep_B - 2 \Re y_f
- \fr{1}{\ga a^2 }
\Bigl[\De \ga b^2 \Re (x_f - \overline x_f)
\nonumber \\
& & \qquad\qquad\qquad\qquad
- 2 \De m \ga_S \ga_L \Im (x_f + \overline x_f)\Bigr]
\quad .
\label{iiid}
\eea
These asymmetries provide insight about the
information available from a complete analysis.
As an example,
in the case where $y_f$, $x_f$, and $\bar x_f$ 
are taken to be negligible,
the first asymmetry gives information about 
indirect CPT violation
while the second reduces to $A_f^\prime = 4 \Re \ep_B$.

We next consider decays into the final states  
of the form $J/\ps K$,
which lie outside the semileptonic-type class.
The physical final states involve 
the linear combinations $K_S$ and $K_L$
given by the kaon equivalent of Eq.\ \rf{iia}.
They are denoted by $\bra{J/\ps K_S}$ and $\bra{J/\ps K_L}$.

The four possible time-dependent decay probabilities,
\bea
P_S(t)
&\equiv &
|\bra{J/\ps K_S}T\ket{B(t)}|^2
\nonumber \\
& = &
(\Re F_{J/\ps})^2 \bigg(
\big(
\half - \Re \ep_B + \Re \de_B 
+ \half(\Re x_{J/\ps} + \Re \overline x_{J/\ps})
\big) \exp (- \ga_S t)
\nonumber \\
& & \qquad\qquad
+ \Big[ \big(
\Re \ep_K + \Re \ep_B + \Re \de_K - \Re \de_B 
\nonumber \\
& & \qquad\qquad\qquad
- \Re y_{J/\ps} - \half\Re (x_{J/\ps} - \overline x_{J/\ps})
\big) \cos \De m t
\nonumber \\
& & \quad\qquad\qquad
- \big(
\Im \ep_K - \Im \ep_B + \Im \de_K + \Im \de_B 
\nonumber \\
& & \qquad\qquad\qquad
+ \half\Im (x_{J/\ps} + \overline x_{J/\ps})
- \fr{\Im F_{J/\ps}}{\Re F_{J/\ps}}
\big) \sin \De m t
\Big] \exp(-\ga t/2)
\bigg)
\quad , \nonumber \\
\overline P_S(t) 
&\equiv &
|\bra{J/\ps K_S}T\ket{\overline B(t)}|^2
\nonumber \\
& = &
P_S(
\ep_K \rightarrow -\ep_K,
\ep_B \rightarrow -\ep_B, 
\de_K \rightarrow -\de_K,
\de_B \rightarrow -\de_B, 
\nonumber \\
& & \qquad\qquad\qquad\qquad
y_{J/\ps} \rightarrow -y_{J/\ps}, 
x_{J/\ps} \leftrightarrow \overline{x}_{J/\ps}^* ,
F_{J/\ps} \rightarrow F_{J/\ps}^*
)
\quad , \nonumber \\
P_L(t) 
&\equiv &
|\bra{J/\ps K_L}T\ket{B(t)}|^2
\nonumber \\
& = &
(\Re F_{J/\ps})^2 \bigg(
\big(
\half - \Re \ep_B - \Re \de_B 
- \half(\Re x_{J/\ps} + \Re \overline x_{J/\ps})
\big) \exp (- \ga_L t)
\nonumber \\
& & \qquad\qquad
+ \Big[ \big(
\Re \ep_K + \Re \ep_B - \Re \de_K + \Re \de_B 
\nonumber \\
& & \qquad\qquad\qquad
- \Re y_{J/\ps} + \half\Re (x_{J/\ps} - \overline x_{J/\ps})
\big) \cos \De m t
\nonumber \\
& & \quad\qquad\qquad
+ \big(
\Im \ep_K - \Im \ep_B - \Im \de_K - \Im \de_B 
\nonumber \\
& & \qquad\qquad\qquad
- \half\Im (x_{J/\ps} + \overline x_{J/\ps})
- \fr{\Im F_{J/\ps}}{\Re F_{J/\ps}}
\big) \sin \De m t
\Big] \exp(-\ga t/2)
\bigg)
\quad , \nonumber \\
\overline P_L(t) 
&\equiv &
|\bra{J/\ps K_L}T\ket{\overline B(t)}|^2
\nonumber \\
& = &
P_L(
\ep_K \rightarrow -\ep_K,
\ep_B \rightarrow -\ep_B, 
\de_K \rightarrow -\de_K,
\de_B \rightarrow -\de_B, 
\nonumber \\
& & \qquad\qquad\qquad\qquad
y_{J/\ps} \rightarrow -y_{J/\ps}, 
x_{J/\ps} \leftrightarrow \overline{x}_{J/\ps}^* ,
F_{J/\ps} \rightarrow F_{J/\ps}^*
)
\quad ,
\label{iiie}
\eea
can be found by a procedure analogous 
to that yielding Eq.\ \rf{iiia}.
The above expressions for $P_S$ and $\overline P_S$
are used as input to our Monte-Carlo simulations
in subsection IIIB.

As in the semileptonic-type case,
insight can be obtained from the fully integrated rates.
The four time-integrated rates corresponding to the
above decay probabilities are
\bea
R_S 
& = & 
\int_{0}^{\infty} dt 
P_S
\nonumber \\
& = & 
(\Re F_{J/\ps})^2 \bigg(
\fr{1} {2\ga_S} - \Big[\fr{1}{\ga_S} - \fr{\ga}{2b^2}\Big] 
(\Re \ep_B - \Re \de_B - \half\Re x_{J/\ps})
\nonumber \\
& & \qquad\qquad
+ \half\Big[\fr{1}{\ga_S} + \fr {\ga}{2b^2}\Big]
\Re \overline x_{J/\ps} 
+ \fr{\ga}{2b^2} (\Re \ep_K + \Re \de_K - \Re y_{J/\ps})
\nonumber \\
& & \qquad\qquad
- \fr{ \De m}{b^2} 
\Big[\Im \ep_K - \Im \ep_B + \Im \de_K + \Im \de_B 
+ \half\Im (x_{J/\ps} + \overline x_{J/\ps})
\nonumber \\
& & \qquad\qquad\qquad\qquad
- \fr{\Im F_{J/\ps}}{\Re F_{J/\ps}}\Big] \bigg)
\quad , \nonumber \\
\overline R_S 
& = & 
\int_{0}^{\infty} dt 
\overline P_S
\nonumber \\
& = & 
R_S(
\ep_K \rightarrow -\ep_K,
\ep_B \rightarrow -\ep_B, 
\de_K \rightarrow -\de_K,
\de_B \rightarrow -\de_B, 
\nonumber \\
& & \qquad\qquad\qquad\qquad\qquad
y_{J/\ps} \rightarrow -y_{J/\ps}, 
x_{J/\ps} \leftrightarrow \overline{x}_{J/\ps}^*
F_{J/\ps} \rightarrow F_{J/\ps}^* ,
)
\quad , \nonumber \\
R_L 
& = & 
\int_{0}^{\infty} dt 
P_L
\nonumber \\
& = & 
(\Re F_{J/\ps})^2 \bigg(
\fr{1} {2\ga_L} 
- \Big[\fr{1}{\ga_L} - \fr{\ga}{2b^2} \Big]
(\Re \ep_B + \Re \de_B + \half\Re x_{J/\ps})
\nonumber \\
& & \qquad\qquad
- \half \Big[\fr{1}{\ga_L} 
+ \fr {\ga}{2b^2}\Big] \Re\overline x_{J/\ps} 
+ \fr{\ga}{2b^2} (\Re \ep_K - \Re \de_K - \Re y_{J/\ps})
\nonumber \\
& & \qquad\qquad
+ \fr{ \De m}{b^2} 
\Big[\Im \ep_K - \Im \ep_B - \Im \de_K - \Im \de_B 
- \half \Im (x_{J/\ps} + \overline x_{J/\ps}) 
\nonumber \\
& & \qquad\qquad\qquad\qquad
- \fr{\Im F_{J/\ps}}{\Re F_{J/\ps}}\Big] \bigg)
\quad , \nonumber \\
\overline R_L 
& = & 
\int_{0}^{\infty} dt 
\overline P_L
\nonumber \\
& = &
R_L(
\ep_K \rightarrow -\ep_K,
\ep_B \rightarrow -\ep_B, 
\de_K \rightarrow -\de_K,
\de_B \rightarrow -\de_B, 
\nonumber \\
& & \qquad\qquad\qquad\qquad\qquad
y_{J/\ps} \rightarrow -y_{J/\ps}, 
x_{J/\ps} \leftrightarrow \overline{x}_{J/\ps}^* ,
F_{J/\ps} \rightarrow F_{J/\ps}^*
)
\quad .
\label{iiif}
\eea
The replacements indicated by a double-headed arrow
involve parameter interchange rather than substitution.

Two theoretically interesting asymmetries can be constructed
from the above decay rates.
One is
\bea
A_S & \equiv & \fr {\overline R_S - R_S}
{\overline R_S + R_S}
\nonumber \\
& = & 2(1 - \fr{\ga \ga_S}{2 b^2}) \Re (\ep_B - \de_B)
+ \fr{\ga \ga_S}{2 b^2} \Re (x_{J/\ps} - \overline x_{J/\ps})
- \fr{\ga \ga_S}{b^2} \Re (\ep_K + \de_K - y_{J/\ps})
\nonumber \\
&&+ \fr{2 \De m \ga_S}{b^2}
\biggl(
\Im \ep_K -\Im \ep_B +\Im \de_K +\Im \de_B
+ \half \Im (x_{J/\ps} + \overline x_{J/\ps}) 
\nonumber \\
&&\qquad\qquad\qquad
- \fr{\Im F_{J/\ps}}{\Re F_{J/\ps}}
\biggr)
\quad .
\label{iiig}
\eea
The other is the analogous quantity for the decays
involving $K_L$:
\bea
A_L &\equiv &\fr {\overline R_L - R_L}
{\overline R_L + R_L}
\nonumber \\
&=& 
A_S (\ga_S \leftrightarrow \ga_L, 
\ep_K \to \ep_K^* ,
\ep_B \to \ep_B^* ,
\de_K \to - \de_K^* , 
\de_B \to - \de_B^* , 
\nonumber \\
& & \qquad\qquad\qquad\qquad
x_{J/\ps} \to -x_{J/\ps}^* ,
\overline x_{J/\ps} \to -\overline x_{J/\ps}^* ,
F_{J/\ps} \to F_{J/\ps}^*
) 
\quad .
\label{iiih}
\eea
Of these two asymmetries,
only the one involving $K_S$ is considered
because detection of the $K_L$ is difficult
with current experimental techniques.

\vglue 0.6cm
{\bf\noindent B. Experiment}
\vglue 0.4cm

In this subsection,
we describe the results of a Monte-Carlo simulation
involving uncorrelated $B$ decays 
observed by a typical detector at 
a high-energy collider.
For definiteness,
we use a detector simulation that smears four-vectors 
according to the resolutions for 
charged-track and neutral-energy measurements
of the OPAL detector~\cite{opal1} at LEP.  

The JETSET7.3 
\cite{jetset} 
event generator 
is used to simulate both $B^0_d$ signal events 
and background events from $Z^0$ decays
into light-quark pairs and into other $B$ hadrons.
As discussed in section II,
we restrict the variables in our simulation
to parameters controlling 
indirect T and indirect CPT violation.
This simplifies many of 
the relevant equations in the previous subsection.
The time-dependent mixing probabilities 
for the signal are sampled from the probability distributions 
given for $B^0 \rightarrow D^{(*)} \ell \nu$ decays 
in Eq.\ \rf{iiia} 
and for $B^0 \rightarrow J/\psi K_S$
in Eq.\ \rf{iiie}. 
The decay length $L$ is found
from the magnitude $p_B$ of the generated momentum 
of the signal $B^0$ meson.

For $B^0 \rightarrow D^{(*)} \ell \nu$ decays,
a typical decay-length resolution $\sigma_L = 200$~$\mu$m is taken
\cite{opallife}.
We assume the boost of the $B^0$ could be estimated 
with techniques using scaled $D \ell$ momenta,
producing a resultant fractional resolution 
$\sigma_{p_B}/p_B = 0.11$
\cite{opalmix}.
Rather than reconstructing vertices 
with the smeared four-vectors, 
the reconstructed proper time $t$
is simply smeared according to 
\beq
\left( \frac{\sigma_t}{t} \right)^2 = 
\left(\frac{\sigma_L}{L} \right)^2 + 
\left(\frac{\sigma_{p_B}}{p_B} \right)^2
\quad .
\label{resol}
\eeq

For the fully exclusive $B^0 \rightarrow J/\psi K_S$ decay, 
an estimated decay-length resolution 
of $\sigma_L = 300$~$\mu$m is used.
The smeared decay length is first found, 
and the momentum reconstructed from the sum 
of the smeared four-vectors of the decay products 
is subsequently used to determine the proper decay time.

We first discuss in detail our analysis of semileptonic decays.
In this case,
a useful quantity for CPT studies
is the time-dependent asymmetry
of decay probabilities given by
\bea
A(f,t) &\equiv &
\fr
{ \overline{P}_{\overline f} (t) -  P_f (t) }
{ \overline{P}_{\overline f} (t) +  P_f (t) }
\nonumber \\
&\approx&
4 \Im \de_B \fr {\sin \De m t} {(1 + \cos \De m t)} 
\quad .
\label{iiij}
\eea
In deriving this result,
we have neglected direct CPT and $\De B = \De Q$ violations
and have approximated $\ga_S \simeq \ga_L$.
The latter implies that $\De \ga$ can be treated as small
over the range of time scales considered.

Another key quantity is the difference 
of time-dependent asymmetries
of decay probabilities,
given by 
\bea
D(f,t) &\equiv &
\fr
{ \overline{P}_{\overline f} (t) -  \overline{P}_f (t) }
{ \overline{P}_{\overline f} (t) +  \overline{P}_f (t) }
-\fr
{P_f (t) -  P_{\overline f} (t) }
{P_f (t) +  P_{\overline f} (t) }
\nonumber \\
&\approx&
-4 \Re \ep_B (1 - \cos \De m t ) + 4 \Im \de_B \sin \De m t
\quad ,
\label{iiija}
\eea
where the same approximations have been made.

To construct these asymmetries experimentally, 
the $b$ flavor of the $B$ meson 
at both the decay and the production times must be determined.
The flavor at the decay time 
is tagged by the charge of the lepton.  
The flavor at the production time can be identified 
using a jet charge technique 
\cite{opalmix}.
It is assumed that this method is
incorrect 20\% of the time
\cite{opalmix,alephmix}.
This mistag probability,
the proper time resolution discussed above,
and the backgrounds to the signal 
all dilute the magnitude of the experimentally observed asymmetry.

The expected levels of various backgrounds are estimated from
previous experimental measurements 
of $\De m_d$~\cite{opalmix,alephmix}.
Approximately 20\% of the selected sample can be expected to be 
combinatorial background without lifetime information,
arising from the primary vertex of the event.  
This fraction of events is folded in
with a distribution in the proper decay time 
consistent with a gaussian distribution of width $\sigma_t$.  
About another 10\% of the selected sample
is expected to arise 
from other genuine $B$-hadron decays with lifetime information
that are misidentified as $B^0 \rightarrow D^{(*)} \ell \nu$.
The decay-time distribution of these events 
is taken to be an exponential consistent 
with the average $B$-hadron lifetime
\cite{pdt}
convoluted with a gaussian of width $\sigma_t$, 
but without the mixing time dependence of Eq.\ \rf{iiia}.

For a perfect detector and with the approximations we have made, 
the form of Eq.\ \rf{iiij} contains a pole structure 
at $\cos \De m t = -1$. 
The observed magnitude of this asymmetry 
would be sensitive 
to the precise knowledge of 
the detector resolution, mistag probabilities,  
and backgrounds. 
We therefore focus instead on 
the difference of asymmetries given in Eq.\ \rf{iiija} 
as an experimental observable
for the semileptonic case.

To test the prospect for measuring or bounding $\Im\de_B$
via semileptonic decays, 
we generated large Monte-Carlo samples of $10^6$ events each 
that included all the above effects for a range of values of
$\Im\de_B$.
For definiteness in this and all other analyses described below
\cite{fn}, 
we set $\ep_B$ to the value
\cite{epsb}
$\ep_B = 0.045$.
For each sample,
the difference in asymmetries 
of Eq.\ \rf{iiija} was constructed.

For $\Im\de_B = 0$, 
we then generated an ensemble of 200 datasets of Monte-Carlo events.
Each such dataset contained
1500 $B^0 \rightarrow D^{(*)} \ell \nu$ decays, 
a statistical sample that
could possibly be reconstructed by combining 
the present data from all the LEP collaborations.  
For the observed $D(f,t)$ distribution 
in each of these simulated datasets,
we performed binned maximum-likelihood fits 
to the expected shape 
for a particular value of $\Im\de_B$.
This determined the average bound that could be set on $\Im\de_B$.  
Subsequently,
we repeated the same process for dataset ensembles 
with $\Im\de_B = 0.1$ and $\Im\de_B = 0.2$.

{}From the ensemble of 200 datasets for the semileptonic case
that were generated using $\Im\de_B=0$, 
the average value of the 95\%-confidence-level (C.L.)
upper bound on $\Im\de_B$ 
that can be placed on a single dataset 
is $\Im\de_B < 0.12$.  
Figure 1 shows the result
from a typical dataset that gives rise to this limit.  
For the dataset illustrated, 
the probability is 25\%  
that the data are from a parent distribution
with $\Im\de_B = 0$. 

Assuming indirect CPT violation indeed exists,
Fig.\ 2 shows typical fits and precisions 
to simulated data generated with
$\Im\de_B = 0.10$ and $\Im\de_B = 0.20$.
In the samples indicated, 
the probabilities that the 
simulated datasets are from
parent distributions with $\Im\de_B = 0$ 
are 0.5\% and 0.1\%,
respectively.

We next turn to an investigation of 
decays into $J/\ps K_S$.
In this case,
a useful quantity is
the time-dependent asymmetry of decay probabilities
given by
\bea
A(J/\ps K_S,t) &\equiv &
\fr
{ \overline{P}_S (t) - P_S (t) }
{ \overline{P}_S (t) + P_S (t) }
\nonumber \\
&\approx&
2 \Re \ep_B - 2 \Re \de_B
-2(\Re \ep_K + \Re \ep_B + \Re \de_K - \Re \de_B) \cos \De m t 
\nonumber \\
&&\qquad\qquad
+2(\Im \ep_K - \Im \ep_B + \Im \de_K + \Im \de_B) \sin \De m t
\quad .
\label{iiik}
\eea
In deriving this expression,
approximations similar to those in Eq.\rf{iiij} 
have been made.
If $\Re \de_B$ and $\Im \de_B$ are to be bounded
at levels greater than known limits on the other parameters, 
then this expression effectively reduces to
\beq
A(J/\ps K_S,t) \approx
-2\Re \de_B (1 - \cos \De m t )
+2\Im \de_B \sin \De m t
\quad .
\label{iiil}
\eeq

We find that the small branching ratio for the hadronic decay
$B^0 \rightarrow J/\ps K_S$ 
and the relatively small reconstruction efficiency
make it unlikely that current LEP data can 
bound CPT effects in this channel.
As an example, 
ALEPH has a sample of only 4 events of this type 
among the $3 \times 10^6$ $Z^0$ decays 
used to measure the $B^0$ lifetime
\cite{alephlife}.  
Samples of $B^0 \rightarrow J/\ps K_S$ 
in uncorrelated $B^0$ events collected by CDF 
at the Tevatron are larger, 
but still insufficient.  
However,
there are more promising prospects for larger samples 
in approximately 1 fb$^{-1}$ of data
to be obtained after the Fermilab main injector 
\cite{main}
begins operation.

For illustrative purposes and using the asymmetry of
Eq.\ \rf{iiil} as the experimental observable
with $\ep_B = 0.045$ as before,
Fig.\ 3a shows the bound that
could be placed on $\Im\de_B$ 
from samples of 500 reconstructed
$B^0 \rightarrow J/\ps K_S$ decays 
including resolutions outlined previously.
The background levels are much smaller 
since this channel is particularly clean.  
Even with these large samples, 
the limits at the 95\% C.L.\ are $\Im\de_B < 0.22$.  
Figure 3b shows the feasibility for
a typical measurement of indirect CPT violation,
assuming an even larger sample of 1500 reconstructed 
$J/\ps K_S$ decays.

\vglue 0.6cm
{\bf\noindent IV. Correlated Systems}
\vglue 0.4cm

{\bf\noindent A. Theory}
\vglue 0.4cm

In this section,
we summarize some key formulae for the situation
where correlated pairs of neutral $B$ mesons are produced.
Since our Monte-Carlo simulations
neglect effects other than from indirect T and CPT violation,
and since the general analysis is already available
in the literature 
\cite{kp2,ck1},
we restrict ourselves here to consideration 
only of those asymmetries
directly useful for the simulations.
A more complete theoretical analysis is
given in refs.\ \cite{kp2,ck1},
where the reader can also find additional details
of some of the derivations below.
For simplicity in what follows,
we take the correlated mesons 
as $B_d$-$\overline{B_d}$ pairs
formed from the decay of the $\Up(4S)$.
Much of the discussion can be extended to the
case of $B_s$ pairs,
along the lines suggested in 
ref.\ \cite{kp2}.

We work in the rest frame of the 
$\Up(4S)$ resonance,
choosing to align 
the $z$-coordinate axis with the
momenta of the $B$ pair.
Since the
the $\Up(4S)$ has $J^{PC}=1^{--}$,
the initial state $\ket{i}$ of the $B$-meson pair
immediately following the decay is
\beq
\ket{i} = \frac 1 {\sqrt 2} 
[\ket{B_S(\z) B_L(-\z)} - \ket{B_L(\z) B_S(-\z)}]
\quad , 
\label{iva}
\eeq
where the argument $(\pm\z )$ indicates 
the direction of the momentum.
To simplify notation,
we label the two mesons by an index $\al = 1,2$
and suppose they decay at times $t_\al$
into final states $\ket{f_\al}$.

The amplitude ${\cal{A}}_{12}(t_1, t_2)$ 
for the decay can be written in terms of the transition amplitudes
$a_{\al S} = \bra{f_\al}T\ket{B_S}$,
$a_{\al L} = \bra{f_\al}T\ket{B_L}$
and their ratios
$\et_\al = a_{\al L}/a_{\al S}$.
The result is
\beq
{\cal{A}}_{12}(t_1, t_2)
=\frac 1 {\sqrt 2} a_{1S}a_{2S}
\bigl (
\eta_2e^{-i(m_St_1+m_Lt_2)-\half (\ga_St_1+\ga_Lt_2)}
-\eta_1e^{-i(m_Lt_1+m_St_2)-\half (\ga_Lt_1+\ga_St_2)}
\bigr ) 
\quad .  
\label{ivb}
\eeq
This expression is the basic input for the
Monte-Carlo simulations in the next two subsections.

The presence of two time variables
implies the existence of several types
of time-integrated rates.
One useful case is obtained by integration
over the linear combination 
\beq
t = t_1 + t_2
\quad .
\eeq
This produces a rate dependent only on 
the time difference,
which reduces some experimental systematics.
Thus,
integrating over $t$ with $\De t$ fixed
yields the once-integrated rate
\bea
I(f_1,f_2, \De t) 
&=& \half \int_{|\De t|}^\infty dt~
|{\cal{A}}_{12}(t_1, t_2)|^2
\nonumber\\
&=& \fr{|a_{1L}a_{2S}|^{2}} {2\ga}
e^{-\ga |\De t| /2}
\Bigl[ e^{-\De \ga \De t /2 }+ 
|r_{21}|^{2}e^{\De \ga \De t/2}
\nonumber\\
&& \quad \qquad \qquad \qquad 
- 2 \Re r_{21} \cos \De m \De t 
- 2 \Im r_{21} \sin \De m \De t 
\Bigr] 
\quad ,
\label{ivd}
\eea
where 
$r_{21} = \et_2/\et_1$.

Other useful once-integrated rates are produced
by integrating over the orthogonal linear combination 
\beq
\De t = t_2 - t_1
\quad 
\eeq
with $t$ held fixed.
The natural possibilities are: 
\bea
J(f_1,f_2,t) 
&=& \half \int_{-t}^t d\De t~
|{\cal{A}}_{12}(t_1, t_2)|^2
\nonumber\\
&=& \half {|a_{1L}a_{2S}|^{2}} 
e^{-\ga t /2}
\Bigl[ 
\fr 1 {\De \ga} (1 + |r_{21}|^2)
(e^{\De\ga t/2} - e^{-\De\ga t/2})
\nonumber\\
&&\qquad \qquad \qquad \qquad \qquad \qquad 
- \fr {2\Re r_{21}}{\De m}
\sin \De m t 
\Bigr] 
\quad ,
\label{ivdb}
\eea
\bea
J^+(f_1,f_2,t) 
&=& \half \int_{0}^t d\De t~
|{\cal{A}}_{12}(t_1, t_2)|^2
\nonumber\\
&=& \half {|a_{1L}a_{2S}|^{2}} 
e^{-\ga t /2}
\Bigl[ 
\fr 1 {\De \ga} 
(1 - e^{-\De\ga t/2}
- |r_{21}|^2 +|r_{21}|^2 e^{\De\ga t/2} )
\nonumber\\
&&\quad \qquad \qquad 
- \fr 1 {\De m}
( \Re r_{21} \sin \De m t +\Im r_{21} (1 - \cos \De m t ))
\Bigr] 
\quad ,
\label{ivdc}
\eea
\bea
J^- (f_1,f_2,t) 
&=& \half \int_{-t}^0 d\De t~
|{\cal{A}}_{12}(t_1, t_2)|^2
\nonumber\\
&=& -e^{-\ga t} J^+ (f_1,f_2,-t )
\quad .
\label{ivdd}
\eea
These two once-integrated rates are used 
for the simulations in the next two subsections.

Integration over the remaining variable 
to produce twice-integrated rates
independent of the times $t_1$ and $t_2$
can be performed with several different integration limits.
If the range is complete,
this procedure produces
the fully time-integrated rate $\Ga(f_1,f_2)$ 
for the double-meson decay into final states $f_1$ and $f_2$.
However,
it is useful to introduce also the partial rates
$\Ga^+(f_1,f_2)$ for which the decay into $f_1$ occurs first 
and $\Ga^-(f_1,f_2)$ for which it occurs second.
This yields the expressions
\bea
\Ga(f_1,f_2) & = & \int_{-\infty}^{\infty} d \De t
\ I(f_1,f_2,\De t) \nonumber \\
& = & \fr 1 {2\ga_S\ga_L} 
\Bigl[ |a_{1S}a_{2L}|^2 + |a_{1L}a_{2S}|^2 - \fr{\ga_S\ga_L}{b^2} 
(a^*_{1S}a^*_{2L}a_{1L}a_{2S} + {\rm c.c.})\Bigr]
\quad ,
\nonumber\\
\Ga^+(f_1,f_2) & = & \int_0^\infty d \De t
\ I(f_1,f_2,\De t) \nonumber \\
& = & \fr 1{2\ga} \Bigl[ 
\fr{|a_{1S}a_{2L}|^2}{\ga_L} + \fr{|a_{1L}a_{2S}|^2}{\ga_S} 
- (\fr{a^*_{1S}a^*_{2L} a_{1L}a_{2S}} 
{\half\ga - i \De m} + {\rm c.c.})
\Bigr]
\quad ,
\nonumber\\
\Ga^-(f_1,f_2) & = & \int_{-\infty}^0 d\De t \ I(f_1,f_2,\De t) 
= \Ga^+(f_1,f_2) 
|_{m_S \leftrightarrow m_L, \ga_S \leftrightarrow \ga_L}
\quad .
\label{ive}
\eea

Given these integrated rates,
we can form asymmetries containing the essential information. 
An asymmetry useful for extracting the
T-violation parameter $\Re \ep_B$ is
\beq
A_{f,\overline f}^{\rm tot} \equiv \fr
{\Ga(f,f) - \Ga(\overline f, \overline f)}
{\Ga(f,f) + \Ga(\overline f, \overline f)} 
= 4 \Re (\ep_B - y_f)
\quad .
\label{ivf}
\eeq
We remark in passing that the incorporation of inclusive rates
permits the extraction of the T-violation parameter $\Re \ep_B$
independently of $\Re y_f$
\cite{ck1}.

To obtain the real part of $\de_B$,
which parametrizes indirect CPT violation,
it is useful to consider double-meson decays 
into a semileptonic-type state $f$ in one channel 
and into $J/\ps K_S$ in the second.
We disregard decays into $J/\ps K_L$ in what
follows because they are difficult to observe experimentally.
The relevant ratio of matrix elements is
\bea
\et_{J/\ps K_S} & \equiv &
\fr{\bra{J/\ps K_S}T\ket{B_L }}
{\bra{J/\ps K_S}T\ket{B_S}}
\nonumber \\
& = & 
\ep_K^* + \ep_B + \de_K^* - \de_B 
- \Re y_{J/\ps} 
-  \half (x_{J/\ps} - \overline x^*_{J/\ps})
+~ i \fr {\Im F_{J/\ps}} {\Re F_{J/\ps}} 
\quad .
\label{ivg}
\eea
Here,
the quantities $\ep_K$ and $\de_K$ are 
measures of indirect T and CPT violation in the kaon-system,
analogous to $\ep_B$ and $\de_B$ in the $B$ system.
We have assumed that $\Im F_{J/\ps}$ is a small quantity,
which is reasonable since it governs direct T violation.
This ratio of matrix elements
enters the rate asymmetry useful for the extraction
of the real part of $\de_B$,
which is 
\bea
A_{f,K_S} & \equiv & 
\fr{\Ga(f,J/\ps K_S) - \Ga(\overline f,J/\ps K_S)} 
{\Ga(f,J/\ps K_S) + \Ga(\overline f,J/\ps K_S)}  
\nonumber \\
& = & 2\Re (\ep_B - y_f - \de_B) 
- \fr{2\ga_S\ga_L}{b^2}\Re\et_{J/\ps K_S}
\quad .
\label{ivh}
\eea
The above formula assumes
that $x_f = \overline x_f$ and $x_{J/\ps} = \overline x_{J/\ps}$,
i.e.,
that violations of the $\De B=\De Q$ rule
are independent of violations of CPT invariance.
The kaon-system parameters and $\ep_B$ are of order
$10^{-3}$ or less,
so this latter asymmetry provides a means of limiting
or measuring $\Re \de_B$ at levels larger than this, 
under the assumption of negligible direct CPT violation.

Once $\Re\de_B$ is known,
the extraction of the imaginary part of $\de_B$
can be performed using another asymmetry
involving double semileptonic decay,
given by
\beq
A_{f,\overline f} \equiv 
\fr {\Ga^+(f,\overline f) - \Ga^-(f,\overline f)} 
{\Ga^+(f,\overline f) + \Ga^-(f,\overline f)} =
4 \fr {b^2 \De \ga \Re\de_B + 2 \De m \ga_S\ga_L \Im\de_B} 
{\ga(b^2 + \ga_S\ga_L)}
\quad .
\label{ivk}
\eeq
In this case,
the derivation assumes
$x_f = \overline x^*_f$,
i.e.,
that any violation of $\De B=\De Q$
is independent of CP violation.

\vglue 0.6cm
{\bf\noindent B. Experiment: unboosted case}
\vglue 0.4cm

In this subsection,
we describe the results of a Monte-Carlo simulation
involving correlated $B$ decays 
observed by a detector at a symmetric $B$ factory,
where the energies
of the colliding electron and positron beams are equal.
For definiteness,
we use a detector simulation equivalent 
to the performance of the upgraded CLEO experiment at CESR,
for which luminosity upgrades should allow the collection 
of large samples of data 
comparable in size and rate 
to those accumulated by the $B$ factories
at SLAC and KEK.

Four-vectors are smeared as described in the previous section, 
except that we use 
appropriate resolutions for charged-track and neutral-energy
measurements for the CLEO~II detector
\cite{cleo1}.  
A decay-length resolution
of $\sigma_L = 150$ $\mu$m 
is taken for the reconstruction of vertices from 
both 
$B^0 \rightarrow D^{(*)} \ell \nu$ decays and 
$B^0 \rightarrow J/\psi K_S$ decays.  
Difficulties with the time resolution
can immediately be anticipated 
since the $B^0$ mesons are produced almost at rest,
which leads to short average decay lengths.  
Simulations confirm that measuring the time dependence
of the decay-time asymmetries 
is infeasible using either the CLEO~II detector 
or even the future CLEO~III~\cite{cleo3} detector.
Indeed, 
the planned silicon-microvertex detectors for CLEO
will primarily be used 
for background rejection rather than for
the measurement of lifetimes.

Despite these difficulties, 
total-rate asymmetries can still be considered. 
Note, for example, 
that the T-violation parameter
$\epsilon_B$ can be measured using the asymmetry of
Eq.\ \rf{ivf}.  
More relevant for our present purposes
is that the time-integrated asymmetry presented in Eq.\ \rf{ivh} 
can be used at a symmetric $B$ factory 
to extract $\Re\de_B$.  

For our simulations,
we take typical CLEO reconstruction efficiencies
and backgrounds for the decay
$B^0 \rightarrow D^{(*)} \ell \nu$ 
(see refs.~\cite{cleop1}) 
and
$B^0 \rightarrow J/\psi K_S$ 
(see ref.~\cite{cleop2}).  
With an integrated luminosity of 100~fb$^{-1}$ 
equivalent to about 54 million correlated 
$B^0$-$\overline{B^0}$ events, 
we estimate reconstruction is possible
for approximately 3200 signal events 
in which one $B^0$ meson decays into
$D^{(*)} \ell \nu$  
while the other decays into $J/\psi K_S$.
The sign of the electric charge 
of the lepton in the semileptonic decay
can be used to determine whether 
an $f$ or $\bar{f}$ state is present.

As discussed in section II,
we restrict the variables in our simulation
to parameters controlling 
indirect T and indirect CPT violation.
Taking $\ep_B = 0.045$ as before, 
the asymmetry of Eq.\ \rf{ivh} 
can be used to set an estimated limit
of $\Re\de_B < 0.03$ at the 95\% C.L.  
Present integrated luminosities
at CLEO of about 2.0~fb$^{-1}$ would give weak bounds 
$\Re\de_B \lsim 0.35$.

\vglue 0.6cm
{\bf\noindent C. Experiment: boosted case}
\vglue 0.4cm

More confidence in the limits or measurements 
of $\de_B$ obtained through its effects
on the observed time evolution of the correlated $B$ system 
can be obtained in asymmetric machines
presently under construction.
The planned $B$ factories at SLAC and KEK will collide
$e^+$ and $e^-$ at different energies.
The result is a moving $\Up (4S)$ that
subsequently decays into $B^0 \overline{B^0}$. 
Each $B^0$ then eventually decays,
producing a secondary vertex that can be reconstructed.  

Since both $B$ mesons are boosted along the beam direction 
and have little transverse momentum, 
the difference in decay points 
is primarily along the beam or $\hat z$ direction.
For a 9~GeV $e^-$ beam colliding with a 3.1~GeV $e^+$ beam 
as being planned for the $B$ factory at SLAC, 
the average separation in $z$ between the two decay vertices 
is approximately 200~$\mu$m.
With a sufficiently sophisticated vertex detector, 
precisions can be attained allowing the measurement 
of the time evolution of the $B^0$-$\overline{B^0}$ system.  
The goal of the $B$ factory is
to measure the time-dependent asymmetry 
in decays to CP eigenstates,
but the effects of CPT violation 
can also be elucidated.

Here,
we describe the results of a Monte-Carlo simulation
involving correlated $B$ decays 
observed by a representative detector at 
a typical asymmetric $B$ factory.
For definiteness,
we use a detector simulation
approximating the future BaBar detector
\cite{tdrb} at SLAC.
With minor changes,
our results should also be valid for
the BELLE experiment at KEK.
As before,
we restrict the variables in our simulation
to parameters controlling 
indirect T and indirect CPT violation.

We again employ smeared four-vectors for these simulations.
The correlated decay times $t_1$ and $t_2$ are sampled
according to the relations given in 
Eq.\ \rf{ivb}, \rf{ivd}, and \rf{ivdb}.
{}From the generated momentum of each signal $B^0$ meson,
two decay lengths $L_1$ and $L_2$ are found. 
We assume
a typical decay-length resolution of $\sigma_L = 60$~$\mu$m
\cite{tdrb}.
The proper decay times $t_1$ and $t_2$ are then extracted for
the $B_S$ and $\overline{B_L}$ decay via
the experimentally accessible channels
$B^0 \rightarrow D^{(*)} \ell \nu$ and/or
$B^0 \rightarrow J/\psi K_S$,
using the techniques described in Section IIIB.

We consider first the case where 
both $B$ mesons decay semileptonically 
into $D^{(*)} \ell \nu$.
The following two useful $t$- or $\De t$-dependent asymmetries
of once-integrated decay probabilities can be obtained:
\bea
A(f,\overline f,\De t) &\equiv &
\fr
{ I(f,\overline f, +|\De t|) - I(f,\overline f, -|\De t|) }
{ I(f,\overline f, +|\De t|) + I(f,\overline f, -|\De t|) }
\nonumber \\
&\approx&
4 \Im \de_B \fr {\sin \De m t} {(1 + \cos \De m t)} 
\quad ,
\label{ivm}
\eea
and
\bea
A(f,\overline f, t) &\equiv &
\fr
{ J^+(f,\overline f, t) - J^+(f,\overline f, t) }
{ J^+(f,\overline f, t) + J^+(f,\overline f, t) }
\nonumber \\
&\approx&
4 \Im \de_B \fr {(1 - \cos \De m t)} {(\De m t +\sin \De m t)} 
\quad .
\label{ivn}
\eea
In these equations, 
direct CPT and $\De B = \De Q$ violations have been neglected,
and $\De \ga t $ and $\De \ga \De t$ have been treated as small.
The latter holds over the time scales considered and
is justified because the lifetime-difference parameter $y$ 
is small.

For a perfect detector and with the approximations made, 
Eq.\ \rf{ivm} contains a pole structure at $\cos \De m t = -1$.
This suggests the measured magnitude of the asymmetry 
is sensitive to precise details of the detector resolution 
and backgrounds.  
For this reason, 
we consider only the asymmetry \rf{ivn} 
as an experimental observable
for the double-semileptonic case. 

The $f$ and $\bar{f}$ states 
are identified from the sign of the electric
charge of the lepton in each semileptonic decay. 
No dilution of the asymmetry is incurred 
through flavor mistagging.
We take background levels and reconstruction efficiencies 
from prior works
\cite{cleop1}.
With an integrated luminosity of 100~fb$^{-1}$, 
we expect a sample size 
of about 2000 reconstructed correlated events 
with both $B^0$ mesons decaying semileptonically.

To examine the feasibility of measuring or bounding $\Im\de_B$
using the double-semileptonic channel,
we followed a Monte-Carlo procedure similar to that 
used in Sec.\ IIIB.
Setting $\ep_B =0.045$ 
and for a range of values of $\Im\de_B$,
we produced large samples of $10^6$ events 
incorporating the above experimental effects.
We then obtained the asymmetry of Eq.\ \rf{ivn} 
for each sample, 
allowing for
effects due to decay-time resolution resulting 
in incorrect determination of which $B$ meson decayed first. 

Assuming first that $\Im\de_B = 0$, 
we created an ensemble of 200 datasets of
Monte-Carlo events,
each containing 2000 decays.  
For the observed $A(f,\bar{f},t)$ distribution 
in each case,
we then performed binned maximum-likelihood fits 
to the expected shape 
for a given value of $\Im \de_B$.
We thereby obtained an average bound 
that can be set on $\Im\de_B$. 

{}From the ensemble of 200 datasets that were generated using
$\Im\de_B=0$, 
the average value of the upper bound
on $\Im\de_B$ that can be placed 
at the 95\% C.L.\ on a single
dataset is $\Im\de_B < 0.08$.  
Figure 4 shows a typical dataset 
that gives rise to this limit.  
For the dataset illustrated, 
the probability is 86\%
that the data are from a parent distribution
with $\Im\de_B = 0$. 

Next,
we address the more interesting case 
for which one $B$ meson decays semileptonically 
while the other decays into $J/\psi K_S$.
This is one of the most attractive $B$-factory channels 
to determine the angle $\be$ 
in the Cabibbo-Kobayashi-Maskawa triangle,
using an asymmetry arising between events 
where the semileptonically tagged $B$ meson decays first 
and those where the decay into $J/\ps K_S$ occurs first
\cite{ckm}.
The exclusive $J/\ps K_S$ decay 
allows for full reconstruction of the $B$ momentum.

For this situation,
we can examine two useful asymmetries:
\bea
A(f,\overline f,J/\ps K_S,\De t) &\equiv &
\fr
{ I(f,J/\ps K_S, \De t) - I(\overline f,J/\ps K_S, \De t) }
{ I(f,J/\ps K_S, \De t) + I(\overline f,J/\ps K_S, \De t) }
\nonumber \\
&\approx&
2 \Re \ep_B - 2 \Re \de_B
\nonumber \\
&&\qquad
-2(\Re \ep_K + \Re \ep_B + \Re \de_K - \Re \de_B) \cos \De m t 
\nonumber \\
&&\qquad
+2(\Im \ep_K - \Im \ep_B + \Im \de_K + \Im \de_B) \sin \De m t
\quad ,
\label{ivo}
\eea
and
\bea
A(f,\overline f,J/\ps K_S,t) &\equiv &
\fr
{ J(f,J/\ps K_S, t) - J(\overline f,J/\ps K_S, t) }
{ J(f,J/\ps K_S, t) + J(\overline f,J/\ps K_S, t) }
\nonumber \\
&\approx&
2 \Re \ep_B - 2 \Re \de_B
\nonumber \\
&&\qquad
-2(\Re \ep_K + \Re \ep_B + \Re \de_K - \Re \de_B) 
\fr{\sin \De m t }{\De mt}
\quad .
\label{ivp}
\eea
In our experimental simulations, 
we use the asymmetry \rf{ivp} 
because it allows a study of $\Re\de_B$ 
independent of $\Im\de_B$.

The procedure we follow
is similar to that for the double-semileptonic case.
Here,
an integrated luminosity of 100~fb$^{-1}$ should result in
2300 reconstructed tagged events~\cite{tdrb}.
The results of
our simulations are illustrated in Fig.\ 5.
If $\Re\de_B = 0$,
an average limit of 
$\Re\de_B < 0.035$ 
at the 95\% C.L.\ can be obtained
with a data sample of this size,
as is shown in Fig.\ 5a.  
Under the assumption that indirect CPT violation 
with $\Re\de_B = 0.20$ indeed exists,
a typical fit and the associated precisions
are illustrated in Fig.\ 5b.
In the sample indicated, 
the probability that the simulated dataset is from a
parent distribution with $\Re\de_B = 0$ 
is less than 0.05\%.  
It may be possible to attain even better
sensitivity by fitting simultaneously to the asymmetry \rf{ivo} 
or by performing a full two-dimensional fit 
to the distributions in $t_1$ and $t_2$.

\vglue 0.6cm
{\bf\noindent V. Summary}
\vglue 0.4cm

In this paper,
we have examined the feasibility of bounding CPT
violation in the neutral-$B$ system.
The type of violation allowed 
might occur in a string scenario
within the context of conventional quantum mechanics.
Our analysis includes the cases 
where the mesons are uncorrelated 
and those where the mesons are correlated
and either unboosted or boosted.

On the purely theoretical side,
we have provided analytical expressions
for both time-dependent and time-integrated decay rates
for all these cases,
allowing for direct and indirect T and CPT violation.
Asymmetries are defined that permit
the extraction of the parameters for indirect CPT violation.
To address the issue of experimental testing  
of these effects,
we have performed Monte-Carlo simulations
to model various plausible experimental situations.
Both data already taken and data likely to be 
available within a few years are considered.
Our treatment incorporates background effects and acceptances
appropriate for the detectors OPAL at CERN,
CLEO at CESR, 
and BaBar at SLAC. 

At present, 
no bounds exist on CPT violation in the $B$ system.
Our analysis suggests that
under reasonable assumptions and
\it with data already available \rm
a bound of order 10\%
can be placed on CPT-violation parameter $\de_B$.
Our simulations also suggest that the CPT reach of
planned experiments is likely to attain the level
of a few percent within the near future.

Given the role of CPT invariance as a fundamental
symmetry of the standard model,
testing it is vital.
If CPT violation were discovered,
it could provide a test of string theory 
and in any event would have far-reaching implications 
for our understanding of nature.
 
\vglue 0.6cm
{\bf\noindent Acknowledgments}
\vglue 0.4cm

We thank Don Colladay,
J.R. Patterson, and Art Snyder
for discussion. 
This work was supported in part
by the United States Department of Energy 
under grant number DE-FG02-91ER40661.

\baselineskip=18pt

\vglue 0.4cm

\newpage
FIGURE CAPTIONS

\bigskip
\noindent
Figure 1.
Simulated data and bound for the uncorrelated case
in the semileptonic channel.
The points with error bars 
represent the typical observed difference of asymmetries $D(f,t)$ 
as a function of time 
for 1500 $B^0 \rightarrow D^{(*)} \ell \nu$ decays
generated in a Monte-Carlo simulation with $\Im\de_B = 0$.
Superimposed as a solid line
is the predicted shape of $D(f,t)$ for 
$\Im\de_B = 0.12$,
which for this sample is the upper bound at the 95\% C.L. 
determined from the difference of likelihood 
in binned likelihood fits.

\bigskip
\noindent
Figure 2.
Simulated data and fits 
for the uncorrelated case in the semileptonic channel for 
(a) $\Im\de_B = 0.1$,
(b) $\Im\de_B = 0.2$.
In each plot,
the points with error bars 
represent the typical observed difference of asymmetries $D(f,t)$ 
as a function of time 
for 1500 $B^0 \rightarrow D^{(*)} \ell \nu$ decays
generated in a Monte-Carlo simulation with 
the indicated value of $\Im\de_B$.
Superimposed as solid lines
are the predicted shapes of $D(f,t)$ 
found from binned likelihood fits.
The fit results and associated precisions are also indicated.

\bigskip
\noindent
Figure 3.
Simulated data for the uncorrelated case 
in the $J/\ps K_S$ channel.

\noindent
(a) The points with error bars 
represent the typical observed asymmetry $A(J/\psi K_S)$ 
as a function of time 
for 500 $B^0 \rightarrow J/\ps K_S$ decays
generated in a Monte-Carlo simulation with $\Im\de_B = 0$.
Superimposed as a solid line
is the predicted shape for 
$\Im\de_B = 0.22$,
which for this sample is the upper bound at the 95\% C.L. 
determined from the difference of likelihood 
in binned likelihood fits.

\noindent
(b) Same but 
for 1500 $B^0 \rightarrow J/\ps K_S$ decays
generated in a Monte-Carlo simulation with $\Im\de_B = 0.2$.
The solid line is the predicted shape.
The fit result and associated precisions are also indicated.

\bigskip
\noindent
Figure 4.
Simulated data and bound in the double-semileptonic channel
at an asymmetric $B$ factory.
Points with error bars 
represent the typical asymmetry $A(f,\bar{f},t)$ 
as a function of total time $t$
obtained in a Monte-Carlo simulation with $\Im\de_B = 0$
using 2000 correlated $B^0 \overline{B^0}$ decays 
with both mesons decaying into $D^{(*)} \ell \nu$.
Superimposed as a solid line
is the predicted shape of $A(f,\bar{f},t)$ for 
$\Im\de_B = 0.08$,
which for this sample is the upper bound at the 95\% C.L. 
determined from the difference of likelihood 
in binned likelihood fits.

\newpage
\noindent
Figure 5.
Simulated data for the semileptonic-$J/\ps K_S$ channel
at an asymmetric $B$ factory.

\noindent
(a) Points with error bars 
represent the typical asymmetry 
$A(f,\bar{f},J/\psi K_S,t)$ 
as a function of total time $t$ 
for a set of 2300 $B^0 \rightarrow J/\ps K_S$,
$\overline{B^0} \rightarrow D^{(*)} \ell \nu$ correlated decays
generated in a Monte-Carlo simulation with $\Re\de_B = 0$.
Superimposed as a solid line
is the predicted shape for 
$\Re\de_B = 0.035$,
which for this sample is the upper bound at the 95\% C.L. 
determined from the difference of likelihood 
in binned likelihood fits.

\noindent
(b) Same but 
for 2300 events 
generated in a Monte-Carlo simulation with $\Re\de_B = 0.2$.
The solid line is the predicted shape.
The fit result and associated precisions are also indicated.


\begin{thebibliography}{xx}
\bibitem{fn}
In a complete experimental analysis,
either a measured value of $\ep_B$ must be used
or fits with varying $\ep_B$
must be performed.
\end{thebibliography}
\end{document}